# Resonant photoemission with circular polarized light on magnetized LSMO

M. C. Richter[1], O. Heckmann[1], B. S. Mun[2], C. S. Fadley[3,4], B. Mercey[5], K. Hricovini[1]

[1]Université de Cergy-Pontoise, 95031 Cergy Pontoise CEDEX, France
[2]Department of Physics and Photon Science, Gwangju Institute of Science and Technology, Gwangju, Republic of Korea (South)
[3]Advanced Light Source, Lawrence Berkeley National Laboratory, Berkeley, CA 94720 USA
[4]Dept. of Physics, University of California Davis, Davis, CA 95616 and Materials Sciences Division, Lawrence Berkeley National Laboratory, Berkeley, CA 94720 USA
[5]CRISMAT, Caen, France

Abstract

We have used resonant photoemission with circular polarized light as a new tool to obtain information about the electronic structure of half-metals. After careful sample surface preparation of $La_{0.7}Sr_{0.3}MnO_3$, we have obtained a dichroic signal for the $L_{2,3}$ absorption of Mn. Working with a magnetized sample and circular polarized light we observe a clear effect of the helicity of light on the position of the resonant Raman-Auger photoelectrons from the *Mn 2p3p3d* decay. These results allow us to achieve a rough estimate of the half-metallic spin half-gap.

PACS numbers: 75.70.-i, 79.60.-i, 79.60.Bm

## I. INTRODUCTION

Half-metal behavior has been theoretically predicted for a variety of Mn compounds. In these systems the majority spin (spin-up) band has metallic character, while the minority (spin-down) band is expected to have a vanishingly small density of states at the Fermi level and hence semiconductor character. For instance, spin-polarized band calculations have predicted half-metallic behavior for the Heusler alloys NiMnSb [1-4], PtMnSb [3,4] and PdMnSb [5], as well as for the oxides $CrO_2$ [6] and $Fe_3O_4$ [7].
Several studies doubt half-metallicity for the $La_{0.7}Sr_{0.3}MnO_3$ perovskite (LSMO) [10]. Different methods of measurement find different values of spin polarization for the valence band [11]. A clear proof of half-metallic behavior was observed in a study on LSMO, in which spin-resolved valence-band photoemission found a clear Fermi cutoff for the majority electrons and an insulating gap for the minority electrons [8]. This work by Park and collaborators is the first spectroscopic confirmation of half-metallicity of LSMO. For years these measurements were not confirmed until recently angle- and spinresolved measurements on in situ prepared samples could show in close collaboration with GGA+U band calculations the importance of k perpendicular broadening [20]. Here halfmetallicity was confirmed inspite the fact that the measured spin polarisation didn't exceed 75% (putting Park's results in question). The calculated gap for the majority spin was found to be about 2.7 eV. In a purely theoretical study, Banach and coworkers found half-metallicity for LSMO via rigid band calculations [9].
In this paper we try to study a possible influence of half-metallicity on resonant photoemission (RPE) with the aim in mind to develop a novel probe of half-metallicity which does not need the experimentally very difficult technique of spin-resolved photoemission and this paper is a first step in this direction.

LSMO at the dopant level we will study exhibits a high value of saturated magnetic moment per Mn atom around 3.7 $\mu_B$ [12] and a Curie temperature of 354 K. Another interesting feature of LSMO material is the metal-to-insulator transition which happens together with the paramagnetic-to-ferromagnetic transition at the Curie temperature [13]. These combined transitions lead to colossal magnetoresistance for this material.

RPE has been widely employed to study the electronic properties of matter. Important information can be derived from the photon-energy dependence of the RPE peaks. For an excitation near an absorption threshold, two different regimes of behavior are observed: in what can be termed the radiationless Raman regime, the kinetic energy of the outgoing electrons follows the incoming photon energy (as is found for non-resonant photoemission), while in the Auger-like regime the electrons are detected at constant kinetic energy. The transition from the Raman to the Auger regime can occur if the electron promoted in the intermediate state delocalizes faster than the core-hole decays, thus not participating to the core-hole recombination process. RPE thus permits investigating the delocalization of the photo-excited electron, as well as other system relaxations on a time-scale given by the core-hole lifetime. In earlier studies, for instance, we have determined the relative time of localization of photon excited *Mn 3d* electrons on samples of very thin (< 0.5 monolayer) Mn layers deposited on Cu (100) by analyzing the Raman/Auger behavior of the *Mn 2p3p3d* decay [14].

If one excites RPE with circular polarized (CP) light on a magnetic sample, one obtains via the well-known Fano effect different populations of spin-up and spin-down states. Here the spin-orbit interaction of the core level involved in the resonant process acts as a spin selector based on the photon helicity. This is the basic origin of x-ray magnetic circular dichroism (XMCD) in x-ray absorption [15]. In this paper, we will use RPE excited with CP radiation to demonstrate the influence of the half-metallicity of LSMO thin films by exciting the *2p → 3d* transition before the absorption edge i.e. in the Raman regime. In this region of photon energy, the photon-excited electron cannot reach an empty *3d*-state without additional energy which it can obtain from an outgoing Auger electron. In this paper, the observed Auger decay process is *Mn 2p3p3d*.

## II. EXPERIMENTS

We have performed the present study at bend-magnet beamline 9.3.2 of the Advanced Light Source (ALS) at the Lawrence Berkeley National Laboratory, which provides left (LCP) and right (RCP) circular-polarized radiation of ~80% polarization. Thin-film samples of LSMO (~1000 Å thickness) deposited on $SrTiO_3$ (STO) were produced by laser ablation in the CRISMAT laboratory, Caen, France. Of importance is the remnant magnetization we obtain with such thin films, which does not exist for bulk LSMO. This aspect is crucial for performing photoemission experiments. The differences between bulk and thin film material are thought to come from strain induced by the STO substrate [16]. The Curie temperature $T_C$ for these thin films was found to be 320K. By optical Kerr measurements, we find a rectangular hysteresis curve showing a weak coercive field of about 30 Oersted at room temperature. However, this type of measurement probes the bulk material and does not give any information about ferromagnetic ordering in the near-surface region that will be probed in RPE. In order to obtain surface sensitivity, we thus also did X-ray magnetic circular dichroism (XMCD) measurements in the total electron yield mode on the $L_{2/3}$ edges of Mn (inset of Fig. 1). The temperature of the sample was about 100 K and thus well below $T_C$. The XMCD signal gives us a measure of the magnetic moment with a probing depth of about 40 Å. The XMCD measurements were done with the sample in remnant magnetization by changing the helicity of the radiation. The light was at grazing incidence along the [010] surface plane direction and a remanent magnetic field was established in the LSMO film along the same [010] surface direction.

Before obtaining any XMCD signal we prepared the sample surface by repeated cycles of soft ion bombarding and subsequent heating in an oxygen atmosphere. Experimental results to be discussed below indicate that we were able to achieve high-quality stoichiometric surfaces at the end of this procedure.

**III. RESULTS AND DISCUSSION**

The absorption spectra shown in the inset of Fig. 1, taken with the two light helicities, and the resulting XMCD spectrum has been corrected in amplitude for the incomplete polarization of light of 80% and for the angle between photon spin and magnetic field of 40°. The maximum amplitude of the XMCD signal at the $L_3$ absorption edge (defined as the difference divided by the sum of absorption spectra obtained with RCP and LCP light) is about 14%. Another experimental proof of obtaining good surface stoichiometry (not shown here) is the metal-to-insulator transition which we observe in valence-band spectra. Heating the sample above $T_C$ leads to a loss of density of states at the Fermi level (the Fermi cutoff), and cooling it down again recovers the metallic Fermi edge.

We have studied the Raman/Auger behavior around the *Mn* $L_3$ resonance of the *Mn 2p3p3d* Auger decay. The *Mn 2p3p3p* decay was not exploitable because of its superposition with the direct *La 4d* photoemission peak for photon energies close to the resonance. In Fig. 1 we show a series of RPE spectra of the *2p3p3d* decay. We have taken spectra in 0.2 eV photon energy steps, but show only a subset of this data in the figure. Based on prior RPE work on a Mn overlayer on Cu [14], we can recognize four primary transitions contained in this decay feature. In the two lower photon energy spectra only three transitions are visible; the forth is probably hidden by the direct *3p* emission. The highest kinetic energy transition represents the direct *3p* photoemission peak, with a dashed line connecting curves linked to this. The dotted line at the bottom of Fig. 1 shows the *3p* photoemission spectrum taken 10 eV below resonance. It is shifted in energy for comparison to our lowest-energy RPE spectrum.
The mixed $Mn^{3+}/Mn^{4+}$ valency in LSMO leaves 3 to 4 electrons in the *3d* band, 3 localized $t_{2g}$ spins in both ionic states, and one delocalized $e_g$ spin for $Mn^{3+}$. For our LSMO composition the number ratio is about $Mn^{3+}/Mn^{4+} = 0.7/0.3 = 2.33$. The coupling between these *3d* electrons (especially the localized $t_{2g}$) and the *3p* hole gives rise to a rich multiplet structure. Multiplet calculations have not yet been done for this system, so we cannot uniquely identify our measured transitions. Nevertheless, we note that the spectra from atomic *Mn* vapor and from thin *Mn* films show as well the four transitions in the *Mn 2p3p3d* decay close to resonance [14]. We will focus here on the most intense transition in the Auger spectra (the lowest kinetic energy feature in the spectra) as measured close to the resonant energy. For this feature we have determined the kinetic energy position at each photon energy. At lower photon energies, i.e. in the pre-edge region for absorption, the direct *Mn 3p* photoemission peak dominates. We have subtracted this peak for photon energies of more than 2 eV below resonance to insure that the position of the maximum of the Auger *2p3p3d* feature could be measured accurately.

In Fig. 2a we show a zoom of the absorption spectra to indicate the photon energy range for which we have plotted the kinetic energy of the maximum of the *2p3p3d* Auger feature as a function of photon energy (Fig. 2b). We have carefully determined the accurate photon energy for each spectrum by acquiring as well the *Pt 4f* direct PE transition from platinum metal in electrical contact with the sample. Fig. 2b shows data obtained with two different radiation polarizations: LCP (solid points) and RCP (open points). In order to compare the spectra leading to the points in Fig. 3 we show the first four spectra (counted from the lowest shown photonenergy) for LCP and RCP excitation. In Fig 3b the direct photoemission peak was subtracted. For higher photonenergies the subtraction did't change any more the position of the maxima. In Fig. 2b, two regimes are clearly identified : classical Auger behavior meaning constant kinetic energy for the higher photon energies, and Raman behavior shown by the line with slope one. The crossing of both lines takes place at the photon energy $E_c$, which we find at about 1.3 eV below the resonance energy (here defined as the maximum of $L_3$ absorption). This energy $E_c$ can be identified with the Fermi-referenced *2p* binding energy, with this being defined as the energy needed to excite a *2p* electron to the first free 3d state above Fermi level [17]. The maximum of absorption and the cross-over from Raman to Auger behavior are marked in both graphs by dotted vertical lines. We observe small oscillations of our experimental points around the straight lines. These oscillations can be tentatively explained by intensity variations of the different multiplet components [18]. This behavior is shown by both measurements, done independently, with LCP light and with RCP light. Furthermore, both light helicities yield graphs that superpose closely in the higher photon-energy regime, providing some indication of the reproducibility of our data. This not only confirms the sample-based origin of these oscillations but it gives us as well confidence in the high accuracy of our determination of the energetic position of the high-binding-energy transition of the *2p3p3d* feature.

Beyond this however, at about 2.5 – 3.2 eV below resonance there is a clear difference between the two polarizations. (Below $\Delta h\nu = -3.2$ eV the signal becomes too weak to be exploitable with certainty.) We can interpret the splitting between LCP and RCP kinetic energies by means of the schematic picture of the density of states for the LSMO half metal shown in Fig. 4. For LCP excitation the minority spin is more strongly excited than for the opposite polarization, for this spin more energy is needed and taken from the out-going electron by the photoexcited electron in order to reach an empty state. For this reason we measure electrons with slightly lower kinetic energy in the average for LCP excitation than for RCP excitation baring in mind that with each polarization both spins are excited. Based on this picture, we can deduce the approximate value of the spin half-gap (2.3 - 3 eV) from the difference of about 0.2 eV between the kinetic energies of the Auger decay features *Mn 2p3p3d* excited respectively with LCP and RCP light. For this simple model simulation we used the fact that excitation with pure RCP light of the $L_3$ transition of a 3d transition metal induces the transition of 62.5% spin up states and 37.5% spin down states, we use our measured absorption intensities at a photon energy in the Raman region where we observe the biggest effect and from this we extract the relative absorption probabilities for spin up and spin down states. We form the sum of two *Mn 2p3p3d* Auger features with an artificially introduced energy splitting of $\Delta S$, which we choose arbitrarily but reasonably to start with giving each feature a weight we have worked out for RHC and LHC excitation respectively. In this way we obtain two different sum-spectra which have different kinetic energies for their maxima. For a $\Delta S$ of 2.3 - 3 eV the final distance between the sum-spectra is 0.18 - 0.21 eV and corresponds to our observation.

The sum of $\Delta S$ and the halfgap measured by spin resolved photoemission is the spin gap $\Delta_1$, which we thus estimate to be 2.9 - 3.6 eV. This value is in agreement with our spin-resolved photoemission studies and close to the result of calculations by Livesay and coworkers [19] who find a spingap of about 2.3 eV.

In conclusion we have shown and interpreted by means of a simple minded model the influence of the helicity of circular polarized excitation on resonant photoemission in the Raman region when working with of a magnetized sample. We hope this work to be a first step to a possible use of RPE with CP light as a detection of halfmetallicity. As well we would be pleased if this work could incite calculations.

**Figure Caption**

Fig. 1: *Mn 2p3p3d* spectra for photon energies close to resonance. The peak of highest kinetic energy corresponds to direct Mn3p photoemission. The bottom spectrum shows a direct *Mn 3p* photoemission spectrum obtained at a photon energy far below resonance, shifted by 10 eV to line up with the bottom RPE spectrum. The inset shows X-ray magnetic circular dichroism (XMCD) in total-electron-yield absorption at the $L_{2,3}$ edges of Mn.

Fig. 2: (a) Expanded XMCD absorption spectrum for the $L_3$ transition region, representing the photon energy region scanned in (b). (b) Each point presents the kinetic energy position of the main transition of the *Mn 2p3p3d* decay as a function of photon energy. The horizontal line corresponds to classical Auger behavior, the line of slope one corresponds to Raman behavior. We indicate by dotted vertical lines the crossing between Raman and Auger behavior and the position of resonance i.e. absorption maximum respectively. The open circles represent RCP excitation and the filled circles LCP excitation.

Fig. 3: Schematic drawing of the density of states for majority and minority spins in LSMO. Indicated here is the spin half- gap ΔS which we estimate from our RPE data.

Fig. 4 : Simulation of the energy splitting observed for the kinetic energies of the Raman *Mn 2p3p3d* features excited with RCP and LCP light respectively. See text for explanations.

Fig. 1

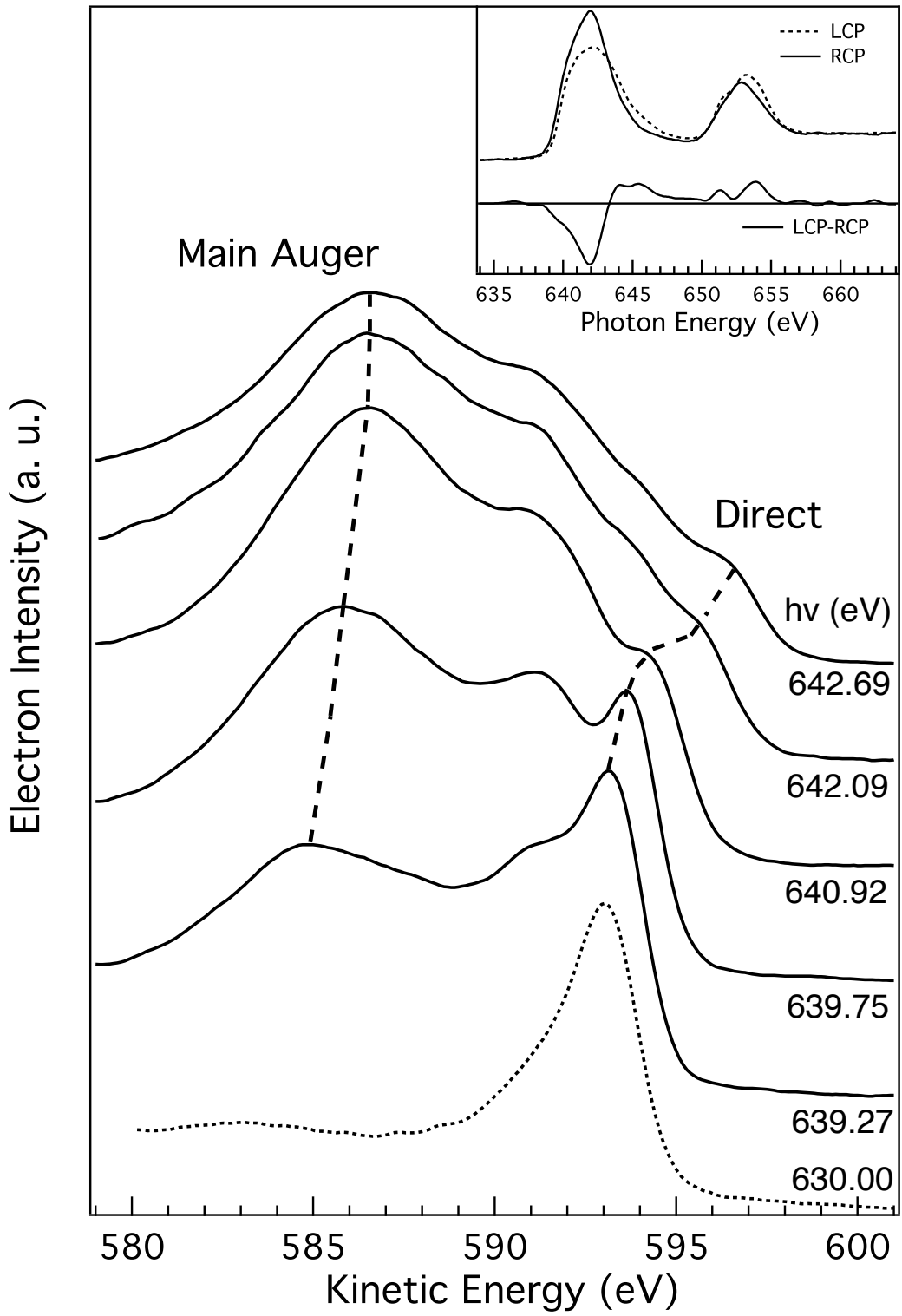

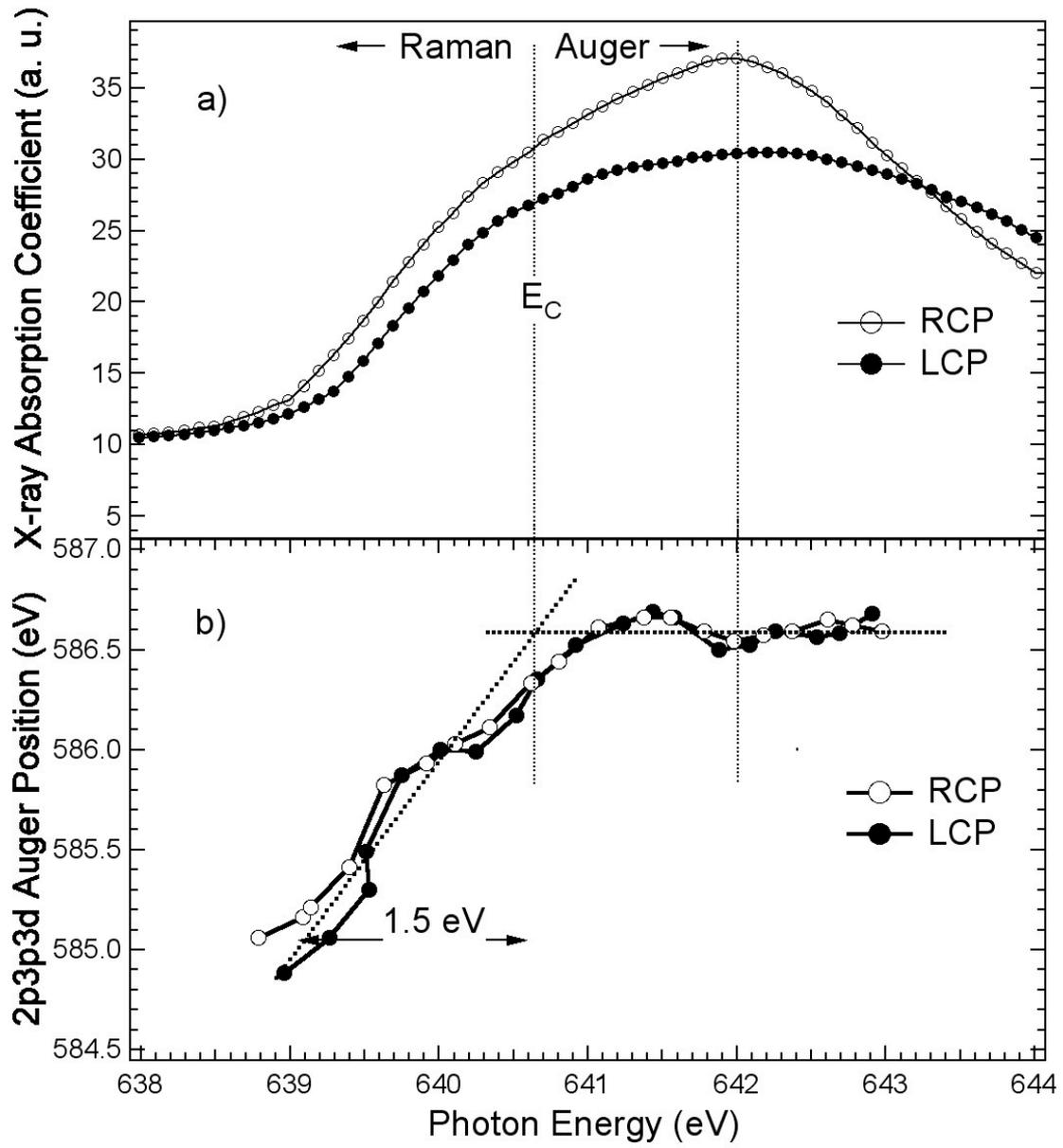
Fig. 2

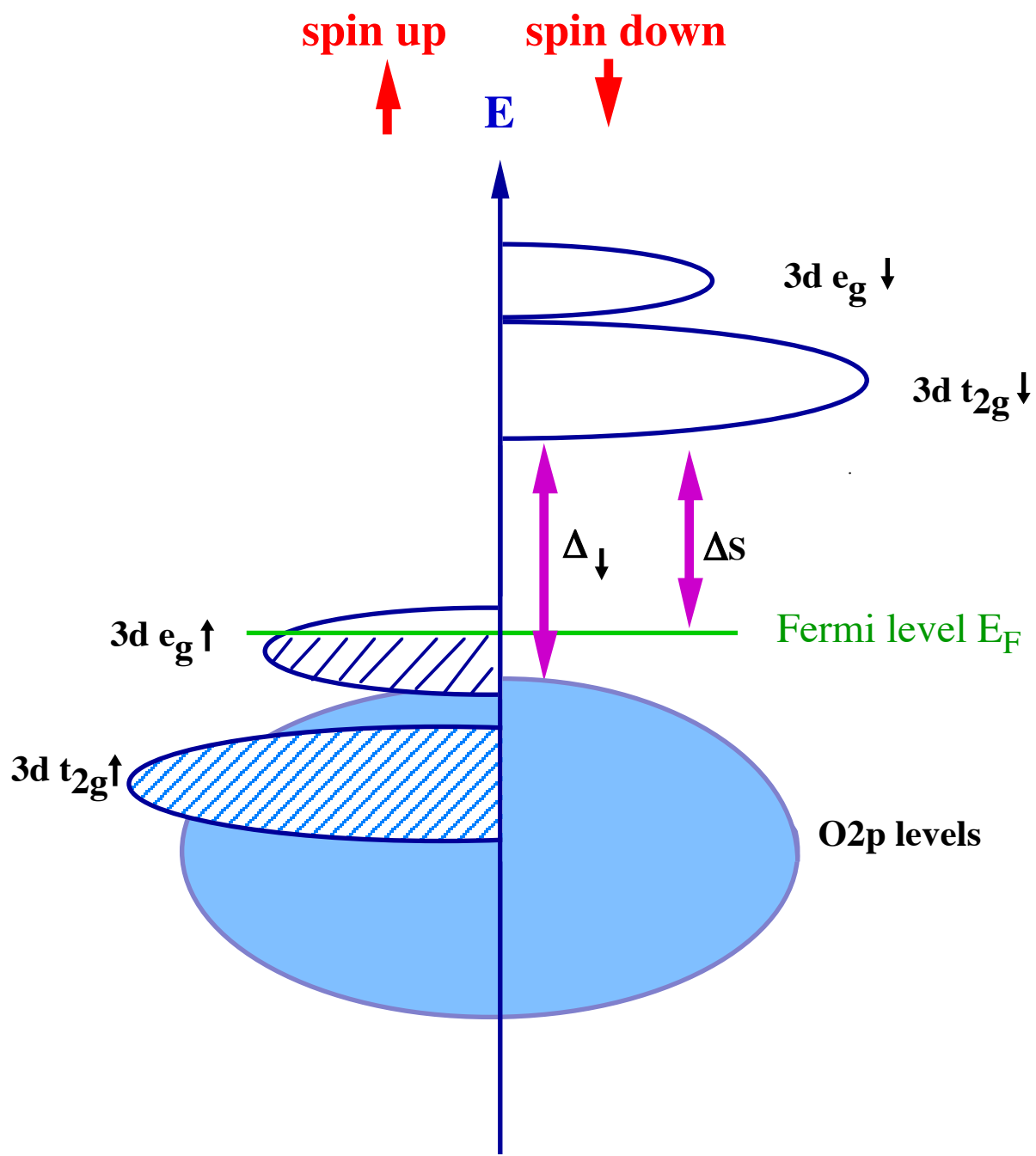

Fig. 3

Fig. 4

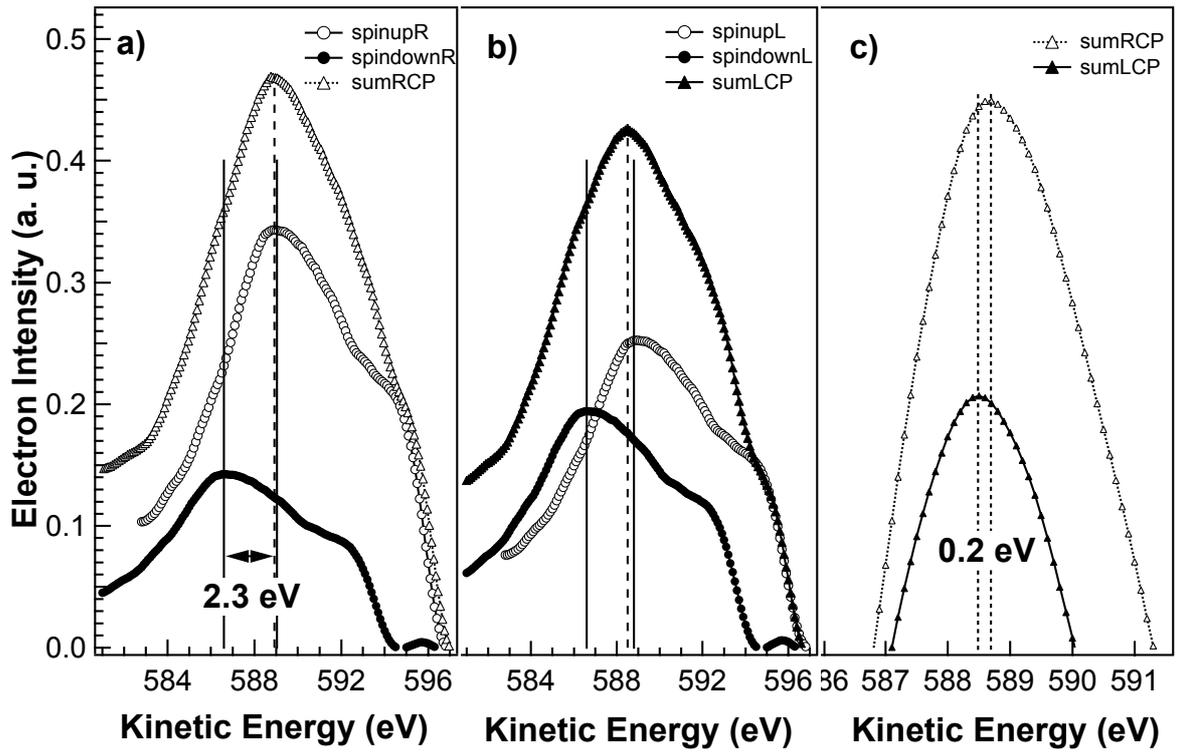